\documentclass[aps,prd,nofootinbib,floatfix,twocolumn]{revtex4}
\usepackage{amssymb}
\usepackage{amsmath}
\usepackage{amsfonts}
\usepackage{graphicx}
\usepackage{bm}

\def\dmsol{\Delta m_s^2}
\def\dmatm{\Delta m_{\rm a}^2}
\def\evolt{{\rm eV}}

\begin{document}
\title{Radiatively induced leptogenesis in a minimal seesaw model}
\author{R. Gonz\'alez Felipe}
\affiliation{Departamento de F\'{\i}sica and Centro de F\'{\i}sica das Interac\c{c}\~{o}es
Fundamentais, Instituto Superior T\'{e}cnico, Av. Rovisco Pais, 1049-001 Lisboa,
Portugal}
\author{F. R. Joaquim}
\affiliation{Departamento de F\'{\i}sica and Centro de F\'{\i}sica das Interac\c{c}\~{o}es
Fundamentais, Instituto Superior T\'{e}cnico, Av. Rovisco Pais, 1049-001 Lisboa,
Portugal}
\author{B. M. Nobre}
\affiliation{Departamento de F\'{\i}sica and Centro de F\'{\i}sica das Interac\c{c}\~{o}es
Fundamentais, Instituto Superior T\'{e}cnico, Av. Rovisco Pais, 1049-001 Lisboa,
Portugal}


\begin{abstract}
We study the possibility that the baryon asymmetry of the universe is generated
in a minimal seesaw scenario where two right-handed Majorana neutrinos with
degenerate masses are added to the standard model particle content. In the
usual framework of thermal leptogenesis, a nonzero $CP$ asymmetry can be
obtained through the mass splitting induced by the running of the heavy
Majorana neutrino masses from their degeneracy scale down to the seesaw scale.
Although, in the light of the present neutrino oscillation data, the produced
baryon asymmetry turns out to be smaller than the experimental value, the
present mechanism could be viable in simple extensions of the standard model.
\end{abstract}

\maketitle

\section{Introduction}

The outstanding advances witnessed by experimental cosmology have brought us to
a new era where an unprecedented precision has been achieved in measuring
several cosmological parameters. Of particular importance is the value of the
baryon-to-photon ratio, which is determined to be~\cite{Spergel:2003cb}
\begin{equation}
\eta_{B}=6.1_{-0.2}^{+0.3}\times10^{-10}\,,\label{BAU}
\end{equation}
from the latest measurements of the Wilkinson Microwave Anisotropy Probe (WMAP)
satellite. Several studies have been aimed at the explanation of this small but
non-vanishing quantity. Among them, leptogenesis~\cite{Fukugita:1986hr} has
become the most attractive mechanism to generate the baryon asymmetry, not only
because of its simplicity, but also due to its possible connection with low
energy neutrino
physics~\cite{Buchmuller:2003gz,Branco:2001pq,Branco:2002xf,Akhmedov:2003dg}.
Indeed, it is conceivable that the physical effects responsible for the
existence of suppressed neutrino masses may simultaneously play a crucial r\^{o}le
in the thermal history of our universe.

From the theoretical point of view, the addition of heavy neutrino singlets to
the standard model particle content is an elegant and economical way to account
for small neutrino masses via the seesaw mechanism. In this framework, the
simplest possibility which can lead to a successful baryogenesis relies on the
out-of-equilibrium $L$-violating decays of heavy Majorana neutrinos (thermal
leptogenesis). The excess of lepton number produced in these decays is then
partially converted into a baryon asymmetry by the ($B+L$)-violating sphaleron
interactions, which are in thermal equilibrium for temperatures $10^2$~GeV
$\lesssim T \lesssim 10^{12}$~GeV.

The heavy Majorana neutrino masses are rather unconstrained in the sense that
they can range from a few TeV up to $10^{16}$~GeV, depending on the model
considered. Moreover, their spectrum may be either hierarchical,
quasi-degenerate or even exactly degenerate. If the heavy Majorana neutrinos
have a hierarchical mass spectrum, a lower bound of $10^8 - 10^9$~GeV can be
inferred on the leptogenesis scale~\cite{Davidson:2002qv}. On the other hand,
the resonant enhancement of the leptonic $CP$ asymmetries through the mixing of
two nearly degenerate heavy Majorana
neutrinos~\cite{Covi:1996wh,Pilaftsis:1997jf} can lower their mass scale up to
TeV energies~\cite{Pilaftsis:1997jf,Pilaftsis:2003gt}.

Due to the proliferation of free parameters in the high energy neutrino sector,
the link between leptogenesis and low energy phenomenology can be only
established in a model dependent way. In the framework of the SM extended with
three heavy Majorana neutrinos one has eighteen parameters in the high-energy
neutrino sector which have to be confronted with the nine low-energy parameters
of the light neutrino mass matrix. A minimal version of the seesaw mechanism,
based on the existence of just two heavy neutrino states, reduces the number of
parameters from eighteen to eleven. But even in this case, additional
assumptions (such as the introduction of texture zeros in the Dirac Yukawa
matrix) are required to completely determine the high energy neutrino sector
from the low energy neutrino observables~\cite{Frampton:2002qc}. The effective
light neutrino mass matrix is in this case defined by seven independent
quantities since one of the neutrinos is predicted to be massless.

In this paper we assume an exact degeneracy of the heavy Majorana neutrinos at
a scale which is higher than their decoupling scale. We show that, within the
minimal seesaw scenario, the heavy neutrino mass splitting induced by
renormalization group effects can lead to appropriate values of the $CP$
asymmetries relevant for the computation of the cosmological baryon asymmetry
in the thermal leptogenesis framework. This provides us with a simple
explanation for the extremely small mass splitting which is typically required
in this context. The compatibility of this framework with the presently
available low-energy neutrino data is also addressed. We show that the solar
and atmospheric neutrino data puts strong constraints on this scenario and
already excludes its minimal version. Since the radiatively induced lepton
asymmetry turns out to be proportional to the charged-lepton $\tau$ Yukawa
coupling, an enhancement of this coupling could be sufficient to obtain the
required baryon asymmetry. This can be achieved, for instance, in simple
extensions of the standard model with more than one Higgs doublet. In this
case, the present framework could easily accommodate the low-energy neutrino
data and lead to a successful leptogenesis.

\section{Seesaw reconstruction}

In the framework of the SM extended with two heavy right-handed neutrino
singlets, and before the decoupling of the heavy Majorana neutrino states, the
leptonic sector is characterized by the Lagrangian
\begin{align}
-{\cal L}  = \bar{\ell}_L Y_\ell\, \ell_{R}\, \phi^0 +\bar{\nu}_L Y_\nu\,
\nu_{R}\, \phi^0 + \frac{1}{2} \nu_{R}^{T} C M_R \nu_{R} + {\rm h.c.}\,,
\label{Lyuk}
\end{align}
where $\ell_{L(R)}\,$ and $\nu_{L(R)}\,$ are the left-handed (right-handed)
charged-lepton and neutrino fields, respectively; $\phi^0$ is the neutral
component of the SM Higgs doublet. The Dirac neutrino and charged-lepton Yukawa
coupling matrices are denoted by $Y_\nu$ and $Y_\ell\,$, while $M_R$ stands for
the $2\times2$ symmetric mass matrix of the heavy right-handed neutrinos. In
the basis where $Y_\ell$ and $M_R$ are diagonal, the neutrino sector is
characterized, in general, by eleven parameters, namely the six moduli and
three phases in the $3\times2$ Dirac neutrino Yukawa coupling $Y_\nu$ plus the
two heavy neutrino masses, $M_1$ and $M_2$. This number may be further reduced
by making assumptions about the structure of
$Y_\nu\,$~\cite{Frampton:2002qc,Raidal:2002xf,Barger:2003gt}.

Let us now assume that at a scale $\Lambda_D$ the heavy neutrinos are
degenerate in mass, i.e. $M_1=M_2 \equiv M$, with $M < \Lambda_D$. We remark
that in the limit of exact heavy Majorana neutrino mass degeneracy, $CP$ is not
necessarily conserved. Indeed, the non-vanishing of the weak-basis
invariant~\cite{delAguila:1996ex,Branco:1998bw}
\begin{equation}
\mathcal{J}_1=M^{-6}\operatorname{Tr}\left[ Y_{\nu} Y_{\nu}^{T}
Y_{\ell}Y_{\ell}^{\dagger} Y_{\nu}^{*}
Y_{\nu}^{\dagger},Y_{\ell}^{*}Y_{\ell}^{T}\right] ^{3},\label{highCPinv}
\end{equation}
which is not proportional to $M_{2}^2-M_{1}^2$, would signal a violation of
$CP$.

It is convenient to define the $3\times3$ seesaw operator at $\Lambda_D\,$,
\begin{align}
\label{seeop} \kappa=Y_\nu^{}\,M_R^{-1}\,Y_\nu^T=\frac{1}{M}Y_\nu^{}
\,Y_\nu^T\,.
\end{align}
An important feature of $\kappa$ is that, at any renormalization scale $\mu$,
one of its eigenvalues is zero leading to a massless neutrino at low energies.
Therefore, this minimal seesaw scenario predicts a hierarchical neutrino mass
spectrum and, as a consequence, a detection of neutrino masses with $m_i >
\sqrt{\dmatm}$ would automatically rule it out, independently of the seesaw
parameters.

For the cases we will be interested in, the decoupling scale of the heavy
neutrino singlets is approximately $M$. Moreover, the RG effects in $Y_\nu$ and
$M_R$ can be neglected for the purposes of the seesaw reconstruction, although
they will be crucial for leptogenesis. Thus, the seesaw operator at the scale
$M$ is well described by Eq.~(\ref{seeop}), apart from an overall factor of
order one.

In order to reconstruct the high energy neutrino sector in terms of the low
energy parameters, we shall consider the following structure for the Dirac
neutrino Yukawa matrix\footnote{For illustration we have chosen $y_2=0$. The
same analysis can be extended to other one-texture-zero Dirac Yukawa
matrices~\cite{Barger:2003gt,prep}.}:
\begin{align}
\label{Ynustruc}
Y_\nu=y_0\left\lgroup
\begin{array}{cc}
  y_1 & 0 \\
  y_3 & y_4 \\
  y_5 & y_6
\end{array}
\right\rgroup\,,
\end{align}
where $y_{1 \dots 6}\,$ are complex and $y_0$ is chosen, without loss of
generality, to be real.

At low energies the effective neutrino mass matrix is given by
\begin{align}
\label{Mnu1}
\mathcal{M}=m_3\,U\,\text{diag}(0,\rho\,e^{i\alpha},1)\,U^T\,,
\end{align}
where $m_3$ is the mass of the heaviest neutrino, $\alpha$ is a Majorana phase
and $\rho \equiv m_2/m_3$. The matrix $U$ is the leptonic mixing matrix which
can be conveniently parametrized in the form
\begin{align}
U  = \left\lgroup
\begin{array}{ccc}
c_{s} c_{r} & s_{s} c_{r} & s_{r}  \\
-s_{s} c_{a} - c_{s} s_{a} s_{r} e^{ i \delta} & c_{s} c_{a} - s_{s} s_{a}
s_{r} e^{ i \delta} &
s_{a} c_{r}e^{i \delta} \\
s_{s} s_{a} - c_{s} c_{a} s_{r} e^{ i \delta} & -c_{s} s_{a} - s_{s} c_{a}
s_{r} e^{ i \delta} & c_{a} c_{r}e^{i \delta}
\end{array}
\right\rgroup, \label{Unu}
\end{align}
where $c_{j} \equiv \cos \theta_{j}\ $, $s_{j} \equiv \sin \theta_{j}\ $ and
$\delta$ is the $CP$-violating Dirac phase.

We will restrict our analysis to the case of neutrinos with normal hierarchy
$0=m_1 < m_2 \ll m_3\,$. In this case
\begin{align}
\label{m2m3r} m_2 = \sqrt{\dmsol}\;,\;m_3 \simeq \sqrt{\dmatm}\;,\; \rho \simeq
\sqrt{\frac{\dmsol}{\dmatm}}\;.
\end{align}
The case of inverted hierarchy can be analyzed in a similar way. In all our
numerical estimates we shall use the best-fit
values~\cite{Gonzalez-Garcia:2004it,Bahcall:2004ut} $\dmsol=8.3 \times
10^{-5}\,\evolt^2$ and $\dmatm=2.2\times 10^{-3}\,\evolt^2$, so that $\rho
\simeq 0.19$. The atmospheric and solar mixing angles are taken as
$\theta_{a}=\pi/4$ and $\tan^2 \theta_{s}=0.37$, respectively. The mixing angle
$\theta_{r}$ is presently constrained to be $|\sin{\theta_{r}}| \lesssim 0.2$
at 95 \% C.L..

The effective neutrino mass operator at the decoupling scale $M$ can be found
by running $\mathcal{M}$ from the electroweak scale $m_Z$ up to $M$. However,
in the case when the spectrum of the light neutrinos is hierarchical, these
effects turn out to be irrelevant. In particular, the mixing angles and the
ratio $\rho$ are stable under the RG evolution. Therefore, one has
\begin{align} \label{seesaw2}
m_3\,U\,\text{diag}(0,\rho\,e^{i\alpha},1)\,U^T\simeq\frac{v^2}{M}Y_\nu^{}\,Y_\nu^T\,,
\end{align}
with $v \simeq 174$~GeV. This relation leads to the following simple
approximation for the reconstructed Dirac neutrino matrix
\begin{align}
H &\equiv Y_{\nu}^{\dagger} Y_{\nu} \simeq \frac{y_0^2}{\sqrt{1+2x^2\cos \alpha+x^4}} \nonumber\\
&\times \left\lgroup
\begin{array}[c]{cc}
\rho+x^2 & x \left[ e^{i \alpha/2}-\rho e^{-i \alpha/2}\right]\\
x \left[ e^{-i \alpha/2}-\rho e^{i \alpha/2}\right] & 1+\rho x^2
\end{array}
\right\rgroup, \label{Hnuapprox}
\end{align}
where
\begin{align} \label{y0}
y_0^2=\frac{M\,\sqrt{\dmatm}}{v^2}\,,\quad
 x=\frac{\tan \theta_r}{\sqrt{\rho}\, s_s} \simeq \frac{s_r}{\sqrt{\rho}\, s_s}\,.
\end{align}

As expected, the assumption of one texture zero allows us to reconstruct the
Dirac neutrino Yukawa matrix (\ref{Ynustruc}) in terms of low energy
observables up to an overall factor which depends on the heavy Majorana
neutrino mass $M$. To determine the latter, additional assumptions would be
required.

\section{Radiatively induced $\bm{CP}$ asymmetries}

A non-zero leptonic asymmetry can be generated if and only if the $CP$-odd
invariant
\begin{align} \label{J2}
    \mathcal{J}_{2} & ={\rm Im\,Tr}\,\left[H M_R^{\dagger} M_R M_R^{\dagger}
    H^T M_R \right] \nonumber\\
   &=M_1 M_2 (M_2^2-M_1^2)\, {\rm Im\,}[H_{12}^2]
\end{align}
does not vanish~\cite{Pilaftsis:1997jf}. This requires not only the heavy
neutrino mass degeneracy to be lifted, but also the nonvanishing of ${\rm
Im\,}[H_{12}^2]$ at the leptogenesis scale $M$. Although the first condition is
easily guaranteed by the running alone of $M_R$ from $\Lambda_D$ to $M$, to
achieve the second condition we must also include the quantum corrections due
to the Dirac neutrino Yukawa matrix $Y_\nu$. Indeed, the running of the
right-handed neutrino mass matrix is governed by the renormalization group
equation (RGE)~\cite{Casas:1999tp,Chankowski:2001mx}
\begin{align}
\label{RGEMR} \frac{d M_R}{dt}=H^T M_R+M_R H\,,\quad
t=\dfrac{\ln\left(\mu/\Lambda_D\right)}{16\,\pi^2}\,.
\end{align}
From this equation, and neglecting the running of $Y_\nu$, one has in the
leading-log approximation $M_R(t) \propto \openone + (H^T+H)\, t$. It is then
possible to show that the trace in Eq.~(\ref{J2}) is a real quantity, which
leads to $\mathcal{J}_{2}=0$. Consequently, the $CP$-violating effects, to
which the $CP$ asymmetries are sensitive, vanish at the decoupling
scale.\footnote{This was first pointed out in Ref.~\cite{Hambye:2004jf}.} Thus,
the corrections coming from the running of $Y_\nu$ from $\Lambda_D$ to $M$ must
be taken into account.

In the basis where $M_R$ is diagonal, and assuming that the charged-lepton
Yukawa matrix $Y_\ell$ is diagonal, the evolution of the right-handed neutrino
mass matrix and the Dirac neutrino Yukawa matrix is given at
one-loop\footnote{In general, the two-loop contribution to the running of $M_R$
enters in the calculation at the same level as the one-loop running of $Y_\nu$.
We have numerically verified that this contribution has a negligible effect on
our results.} by~\cite{Casas:1999tp,Chankowski:2001mx}
\begin{align}
\label{RGEMRdiag}
\frac{d M_i}{dt}&=2 M_i\, H_{ii}\,,\nonumber\\
\frac{d Y_\nu}{dt}&=\left[ {\cal T}- \frac{3}{4} g_Y^2 - \frac{9}{4} g_2^2
-\frac{3}{2} \left(Y_\ell Y_\ell^\dagger-Y_\nu Y_\nu^\dagger\right) \right]
Y_\nu + Y_\nu R\,,
\end{align}
where ${\cal T} = 3 \text{Tr}(Y_u Y_u^\dagger)+3 \text{Tr}(Y_d
Y_d^\dagger)+\text{Tr}(Y_\ell Y_\ell^\dagger)+\text{Tr}(Y_\nu Y_\nu^\dagger)$;
$Y_{u,d}$ are the up-quark and down-quark Yukawa matrices and $g_{Y,2}$ are the
gauge couplings. The matrix $R$ is antihermitian with
\begin{align} \label{Rmatrix}
R_{11}&= R_{22}=0, \quad R_{21} = -\; R_{12}^\ast\,, \nonumber\\
R_{12}&=\frac{2+\delta_N}{\delta_N}\, \text{Re}\,(H_{12})+
i\frac{\delta_N}{2+\delta_N}\, \text{Im}\,(H_{12})\,.
\end{align}
The parameter
\begin{align}
\label{deltaN1} \delta_N \equiv \frac{M_2}{M_1}-1\,,
\end{align}
quantifies the degree of degeneracy between $M_1$ and $M_2$.

An important feature of the system (\ref{RGEMRdiag})-(\ref{Rmatrix}) is the
fact that, if the heavy Majorana neutrinos are exactly degenerate in mass
($\delta_N=0$) at a given scale $\Lambda_D$, then the second RGE in
(\ref{RGEMRdiag}) becomes singular, unless one imposes $\text{Re}\,(H_{12})=0$.
The latter condition can always be satisfied by performing an additional
rotation in the Dirac Yukawa matrix,
\begin{align}
Y_{\nu}^\prime=Y_\nu O\,,\quad O = \left(%
\begin{array}{cc}
  \cos \theta & \sin \theta \\
  -\sin \theta & \cos \theta \\
\end{array}%
\right)\,,
\end{align}
with the rotation angle $\theta$ determined by the relation
\begin{align} \label{tan2theta}
\tan 2\theta = \frac{2\text{Re}\,(H_{12})}{H_{22}-H_{11}}\,.
\end{align}
The rotated matrix $H^\prime=O^\dagger H O$ becomes
\begin{align} \label{Hnurotated}
    H^\prime =\left(%
\begin{array}{cc}
  H_{11}-\Delta & i\, \text{Im}\,(H_{12}) \\
  -i\, \text{Im}\,(H_{12}) & H_{22}+\Delta \\
\end{array}%
\right)\,,
\end{align}
where $\Delta \equiv \tan \theta\, \text{Re}\,(H_{12})$.

According to Eqs.~(\ref{RGEMRdiag}), the running of the mass splitting
parameter $\delta_N$ is determined by the equation
\begin{align}
\label{deltaNrge} \frac{d \delta_N}{dt}=2 (1+\delta_N)
(H_{22}^\prime-H_{11}^\prime)\,.
\end{align}
Given that Eqs.~(\ref{Hnuapprox}), (\ref{tan2theta}) and (\ref{Hnurotated})
imply $H_{22}^\prime-H_{11}^\prime \simeq y_0^2 (1-\rho)$, with $y_0$ defined
in Eq.~(\ref{y0}), the radiatively induced mass splitting at the decoupling
scale $M$ will be approximatively given by
\begin{align}
\label{deltaNM} \delta_N \simeq
\frac{y_0^2}{8\pi^2}\,(1-\rho)\ln\left(\Lambda_D/M\right)\,.
\end{align}
Moreover, even if at the degeneracy scale $\Lambda_D$ one has
$\text{Re}\,(H_{12}^\prime)=0$, a nonvanishing real part will be generated by
quantum corrections. The latter are easily estimated from
Eqs.~(\ref{RGEMRdiag}), and one finds
\begin{align}
    \text{Re}\,(H_{12}^\prime) \simeq \frac{3y_0^2}{32\pi^2}
    \sqrt{\rho}\, y_\tau^2\,
    \text{Re}\,(e^{i\alpha/2} U_{32} U^\ast_{33})\ln\left(\Lambda_D/M\right)\,,
\end{align}
where $y_\tau$ is the $\tau$ Yukawa coupling and $U$ is the neutrino mixing
matrix defined in Eq.~(\ref{Unu}).

The crucial ingredient for the computation of the baryon asymmetry in a
leptogenesis scenario are the $CP$ asymmetries generated by the interference of
the one-loop and tree-level heavy Majorana decay diagrams. In the case when the
heavy Majorana neutrinos are nearly degenerate in mass, these asymmetries are
approximately given by~\cite{Pilaftsis:2003gt}
\begin{align}\label{e12d2}
\varepsilon_j&=\frac{{\rm
Im}\,[\,H^{\prime\,2}_{21}]}{16\,\pi\,\delta_N\,H^{\prime}_{jj}}\left(1 +
\frac{\Gamma_i^2}{4M^2\delta_N^2}\right)^{-1}\,,\quad j=1,2\,,
\end{align}
where
\begin{align}
\Gamma_{i}=\frac{H^{\prime}_{ii} M_i}{8\pi}
\end{align}
are the tree-level decay widths\footnote{One may wonder whether finite
temperature effects could modify the perturbative result. Since the thermally
induced mass splitting is approximately given by $M_2(T)-M_1(T) \lesssim
\frac{1}{16}{\rm Re} (H_\nu^{\prime})_{21} T^2/M$~\cite{Pilaftsis:2003gt}, at
$T \simeq M$ this contribution is very suppressed with respect to the
zero-temperature mass splitting $\delta_N$.}. We notice that the expressions
(\ref{e12d2}) for the $CP$ asymmetries $\varepsilon_i$ exhibit the expected
enhancement due to the mixing of two nearly degenerate heavy Majorana
neutrinos~\cite{Covi:1996wh,Pilaftsis:1997jf}.

\begin{figure}[t]
\begin{center}
\includegraphics[width=8.5cm]{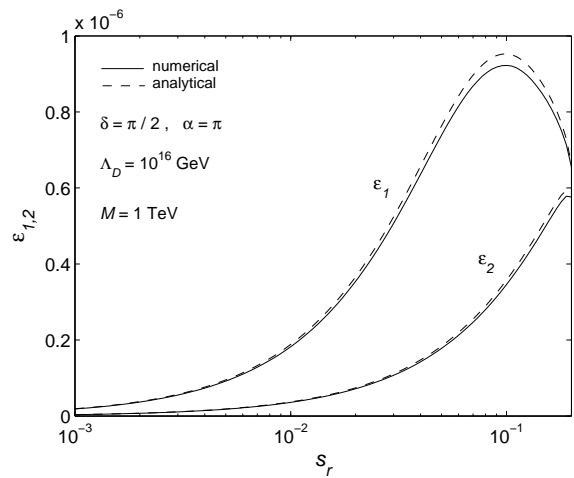}
\caption{Comparison of the analytical and numerical result for the $CP$
asymmetries $\varepsilon_1$ and $\varepsilon_2$ as functions of $s_r =
\sin\theta_{r}$. The analytical curves are defined in Eqs.~(\ref{CPapp}), while
the numerical ones were obtained by integrating the full set of RGEs from the
heavy neutrino degeneracy scale $\Lambda_D$ to the decoupling scale $M$.}
\label{fig1}
\end{center}
\end{figure}

In terms of the low-energy neutrino parameters, we obtain
\begin{align}
\label{CPapp} \varepsilon_1 &\simeq -\frac{3y_\tau^2}{64\pi} \frac{\sqrt{\rho}
\,(1+\rho)x}{(1-\rho)(\rho+x^2-\Delta)} \nonumber\\
&\times \sin (\alpha/2) \left[c_s \cos (\delta-\alpha/2) + \sqrt{\rho}\,
s_s^2\, x \cos (\alpha/2)\right]\,, \nonumber\\
\nonumber\\
\frac{\varepsilon_2}{\varepsilon_1} &\simeq \frac{\rho+x^2-\Delta}{1+\rho
x^2+\Delta}\,,\\ \nonumber\\
\Delta &=
\frac{1}{2}(1-\rho)\left[-1+x^2+\sqrt{1+2x^2\cos \alpha+x^4} \right]\,.
\nonumber
\end{align}
Taking for instance $\alpha \simeq \pi$ and $\delta \simeq \pi/2$, the $CP$
asymmetry $\varepsilon_1$ reaches its maximum value for $x = \sqrt{\rho}\,$, as
can be readily seen from Eqs.~(\ref{CPapp}). This corresponds to $s_r = \rho
s_s \simeq 0.1$ and
\begin{align}
    \varepsilon_1^{\text{max}} \simeq -\frac{3y_\tau^2\,c_s}{128\pi} \frac{
(1+\rho)}{(1-\rho)} \simeq -10^{-6}\,.
\end{align}

The accuracy of these approximate expressions is shown in Fig.~\ref{fig1},
where the $CP$ asymmetries $\varepsilon_i$ are plotted as functions of $s_r$
taking $\Lambda_D=10^{16}$~GeV, $M=1$~TeV, $\delta=\pi/2$, $\alpha=\pi$ and
assuming $y_\tau =0.01$ in the analytical estimates. The solid lines correspond
to the full numerical integration of the RGE system, while the dashed ones
refer to the approximate values given in Eqs.~(\ref{CPapp}). The comparison of
the curves shows that, for values of $s_r \lesssim 0.1$, $\varepsilon_2$ is
suppressed with respect to $\varepsilon_1$, in accordance with
Eqs.~(\ref{CPapp}).
\begin{figure}[t]
\begin{center}
\includegraphics[width=8.5cm]{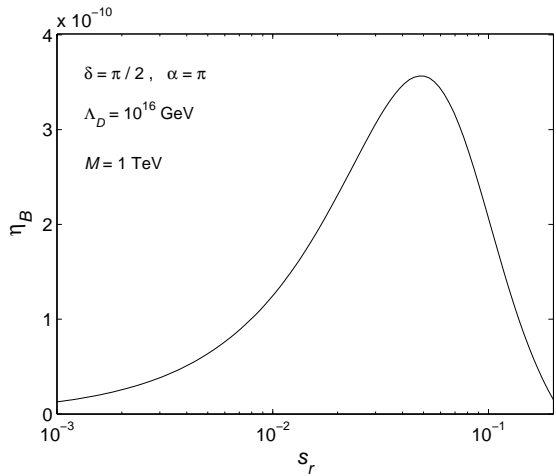}
\caption{Numerical result for the baryon-to-photon ratio $\eta_B$ as a function
of $s_r = \sin\theta_{r}$, obtained by integrating the full set of RGEs from
the heavy neutrino degeneracy scale $\Lambda_D$ to the decoupling scale $M$.}
\label{fig2}
\end{center}
\end{figure}

\section{Cosmological baryon asymmetry}

The out-of-equilibrium Majorana decays are controlled by the parameters
\begin{align} \label{Kfactors}
K_i=\frac{\Gamma_{i}}{H(T=M_i)}\,,
\end{align}
where $H(T)=1.66 g_\ast^{1/2} T^2/M_P$ is the Hubble parameter, $g_\ast \simeq
107$ is the number of relativistic degrees of freedom and $M_P = 1.2 \times
10^{19}$~GeV is the Planck mass. Assuming that the entropy remains constant
while the universe cools down from $T \simeq M$ to the recombination epoch, one
can estimate the baryon-to-photon ratio as $ \eta_B \simeq -10^{-2}\,
(d_1\,\varepsilon_1 + d_2\,\varepsilon_2)\,$, where $d_i \leq 1$ are dilution
factors which account for the washout effects. Since from
Eqs.~(\ref{Hnuapprox}), (\ref{CPapp}) and (\ref{Kfactors}) it follows that
$|\varepsilon_2/K_2| \ll |\varepsilon_1/K_1|$, one expects
\begin{align}
\label{BAUapp} \eta_B \simeq -10^{-2}\, d_1\,\varepsilon_1\,.
\end{align}

We also note that $K_1$ is independent of $M$ and approximately given by
\begin{align} \label{K1app}
    K_1 \simeq \frac{44\, (\rho+x^2-\Delta)}{\sqrt{1+2x^2\cos \alpha+x^4}}\,.
\end{align}
In this washout regime, a simple estimate of the dilution factor $d_1$ can be
obtained from the fit $d_1 \simeq 0.6\, [\ln
(K_1/2)]^{-0.6}/K_1$~\cite{Kolb:vq}. For $\alpha \simeq \pi,\, \delta \simeq
\pi/2$, the maximal value of the baryon asymmetry is then attained for $x
\simeq \sqrt{3\rho/(1+\rho)}/3 \simeq 0.23$ and $s_r \simeq 0.05$. From
Eq.~(\ref{CPapp}) we find $\varepsilon_1 \simeq -8\times10^{-7}$. Moreover,
since Eq.~(\ref{K1app}) implies $K_1 \simeq 11$, then $d_1 \simeq 4 \times
10^{-2}$, using the above fit. Inserting these values into Eq.~(\ref{BAUapp})
one finally gets
\begin{align}
\eta_B^{\text{max}} \simeq 3 \times 10^{-10}\,,
\end{align}
which is by a factor two smaller than the observed baryon asymmetry [cf.
Eq.~(\ref{BAU})]. It is worth noticing that this result is weakly dependent
(through the renormalization effects on $y_\tau^2$) on the heavy Majorana
neutrino mass scale $M$, as can be seen from the approximate expressions given
in Eqs.~(\ref{CPapp}). In Fig.~\ref{fig2} we present the numerical estimate for
the baryon-to-photon ratio $\eta_B$ as a function of $s_r$.

We remark that a more accurate computation of the dilution factors requires the
solution of the full set of Boltzmann equations. This question has been
readdressed in recent works, where the effects on leptogenesis due to the
$\Delta L = 1$ processes involving gauge
bosons~\cite{Pilaftsis:2003gt,Giudice:2003jh} and thermal corrections at high
temperature~\cite{Giudice:2003jh} have been discussed. However, since the
present neutrino data fixes the decay parameters $K_i$ to be $K_i \gg 1$, the
above effects turn out to be negligible and do not alter significantly the
final value of the baryon asymmetry.

\section{Conclusions}

We have considered a minimal seesaw scenario with two heavy Majorana neutrinos
which are degenerate in mass at a scale higher than their decoupling scale. We
have shown that the heavy neutrino mass splitting, induced by radiative
corrections, leads to non-zero $CP$ asymmetries, and subsequently, to thermal
leptogenesis even for heavy Majorana neutrino masses as low as a few TeV.
However, we have seen that in the minimal extension of the SM with the addition
of only two heavy Majorana neutrinos, it seems difficult to reconcile the
present solar and atmospheric neutrino data with the observed cosmological
baryon asymmetry. On the other hand, the analytical expressions (\ref{CPapp})
already suggest that these constraints can be easily relaxed, for e.g., in
extensions of the SM with more than one Higgs doublet, since in this case the
charged-lepton Yukawa couplings can be substantially enhanced.

We also remark that the results obtained here depend on the specific structure
of the Dirac neutrino Yukawa matrix. In particular, it would be interesting to
disentangle the texture considered here from other seesaw textures which can
lead to the same low-energy observables. A more general analysis regarding this
issue will be presented elsewhere~\cite{prep}. It would also be desirable to
find a symmetry which naturally leads to degenerate heavy Majorana neutrinos at
high scales and, simultaneously, to the required Dirac neutrino Yukawa
couplings.

Although low-energy (TeV-scale) leptogenesis is viable in the present
framework, it requires tiny Dirac neutrino Yukawa couplings of the order of
$10^{-6}~-~10^{-7}$, which makes it difficult to be testable at future
colliders. On the other hand, the fact that the final baryon asymmetry weakly
depends on the heavy Majorana neutrino scale allows one to lower this scale
below the typical values ($\sim 10^{8-9}$~GeV) required in the standard thermal
leptogenesis scenarios. Thus, the potential problems related with the
overproduction of relic abundances, which could jeopardize the successful
nucleosynthesis predictions, are avoided.

\emph{Note added}: While this work was being revised,
Ref.~\cite{Turzynski:2004xy} appeared, where it is shown that in the
supersymmetric minimal seesaw model the present mechanism is indeed viable and
thermal leptogenesis is successful.

\begin{acknowledgments}
We are grateful to G.~C.~Branco for valuable discussions and to T. Hambye and
M. Raidal for useful comments. The work of R.G.F., F.R.J. and B.M.N. was
supported by \emph{Funda\c{c}\~{a}o para a Ci\^{e}ncia e a Tecnologia} (FCT, Portugal)
under the grants SFRH/BPD/1549/2000, SFRH/BPD/14473/2003 and SFRH/BD/995/2000,
respectively.
\end{acknowledgments}


\begin{thebibliography}{99}

\bibitem{Spergel:2003cb}
D.~N.~Spergel \textit{et al.},
Astrophys.\ J.\ Suppl.\ \textbf{148}, 175 (2003).

\bibitem{Fukugita:1986hr}
M.~Fukugita and T.~Yanagida,
Phys.\ Lett.\ B \textbf{174}, 45 (1986).

\bibitem{Buchmuller:2003gz}
W.~Buchm\"{u}ller, P.~Di Bari and M.~Pl\"{u}macher,
Nucl.\ Phys.\ B \textbf{665}, 445 (2003).

\bibitem{Branco:2001pq}
G.~C.~Branco, T.~Morozumi, B.~M.~Nobre and M.~N.~Rebelo,
Nucl.\ Phys.\ B \textbf{617}, 475 (2001);
G.~C.~Branco, R.~Gonz\'{a}lez Felipe, F.~R.~Joaquim and M.~N.~Rebelo,
Nucl.\ Phys.\ B \textbf{640}, 202 (2002).

\bibitem{Branco:2002xf}
G.~C.~Branco, R.~Gonz\'{a}lez Felipe, F.~R.~Joaquim, I.~Masina, M.~N.~Rebelo
and C.~A.~Savoy,
Phys.\ Rev.\ D \textbf{67}, 073025 (2003).

\bibitem{Akhmedov:2003dg}
E.~K.~Akhmedov, M.~Frigerio and A.~Y.~Smirnov,
JHEP \textbf{0309}, 021 (2003).

\bibitem{Davidson:2002qv}
S.~Davidson and A.~Ibarra,
Phys.\ Lett.\ B {\bf 535}, 25 (2002);
W.~Buchm\"{u}ller, P.~Di Bari and M.~Plumacher,
Nucl.\ Phys.\ B {\bf 643}, 367 (2002).

\bibitem{Covi:1996wh}
L.~Covi, E.~Roulet and F.~Vissani,
Phys.\ Lett.\ B {\bf 384}, 169 (1996).

\bibitem{Pilaftsis:1997jf}
A.~Pilaftsis,
Phys.\ Rev.\ D {\bf 56}, 5431 (1997).

\bibitem{Pilaftsis:2003gt}
A.~Pilaftsis and T.~E.~J.~Underwood,
Nucl.\ Phys.\ B {\bf 692}, 303 (2004).


\bibitem{Frampton:2002qc}
P.~H.~Frampton, S.~L.~Glashow and T.~Yanagida,
Phys.\ Lett.\ B {\bf 548}, 119 (2002).

\bibitem{Raidal:2002xf}
M.~Raidal and A.~Strumia,
Phys.\ Lett.\ B {\bf 553}, 72 (2003).

\bibitem{Barger:2003gt}
V.~Barger, D.~A.~Dicus, H.~J.~He and T.~j.~Li,
Phys.\ Lett.\ B {\bf 583}, 173 (2004).

\bibitem{prep}
G.~C.~Branco, R. Gonz\'alez Felipe, F. R. Joaquim and B. M. Nobre, in
preparation.

\bibitem{Bahcall:2004ut}
J.~N.~Bahcall, M.~C.~Gonzalez-Garcia and C.~Pena-Garay,
hep-ph/0406294;
A.~Bandyopadhyay, S.~Choubey, S.~Goswami, S.~T.~Petcov and D.~P.~Roy,
hep-ph/0406328.

\bibitem{Gonzalez-Garcia:2004it}
M.~C.~Gonzalez-Garcia and M.~Maltoni,
hep-ph/0406056.

\bibitem{delAguila:1996ex}
F.~del Aguila, J.~A.~Aguilar-Saavedra and M.~Zralek,
Comput.\ Phys.\ Commun.\  {\bf 100}, 231 (1997).

\bibitem{Branco:1998bw}
G.~C.~Branco, M.~N.~Rebelo and J.~I.~Silva-Marcos,
Phys.\ Rev.\ Lett.\ \textbf{82}, 683 (1999).

\bibitem{Casas:1999tp}
J.~A.~Casas, J.~R.~Espinosa, A.~Ibarra and I.~Navarro,
Nucl.\ Phys.\ B {\bf 556}, 3 (1999).

\bibitem{Chankowski:2001mx}
P.~H.~Chankowski and S.~Pokorski,
Int.\ J.\ Mod.\ Phys.\ A {\bf 17}, 575 (2002).

\bibitem{Hambye:2004jf}
T.~Hambye, J.~March-Russell and S.~M.~West,
JHEP {\bf 0407}, 070 (2004).


\bibitem{Kolb:vq}
E.~W.~Kolb and M.~S.~Turner, \textit{The Early Universe},
(Addison-Wesley, Reading, MA, 1990).

\bibitem{Giudice:2003jh}
G.~F.~Giudice, A.~Notari, M.~Raidal, A.~Riotto and A.~Strumia,
Nucl.\ Phys.\ B {\bf 685}, 89 (2004).

\bibitem{Turzynski:2004xy}
K.~Turzynski,
Phys.\ Lett.\ B {\bf 589}, 135 (2004).

\end{thebibliography}
\end{document}